\documentclass[conference]{IEEEtran}
\IEEEoverridecommandlockouts
\usepackage{cite}
\usepackage{amsmath,amssymb,amsfonts}
\usepackage{algorithmic}
\usepackage{graphicx}
\usepackage{textcomp}
\usepackage{xcolor}
\def\BibTeX{{\rm B\kern-.05em{\sc i\kern-.025em b}\kern-.08em
    T\kern-.1667em\lower.7ex\hbox{E}\kern-.125emX}}

\usepackage{array,multirow}
\usepackage{moresize}
\usepackage{amsmath}
\usepackage{amsfonts}
\usepackage{amssymb}
\usepackage{mathtools}
\usepackage{wrapfig}
\usepackage{booktabs}

\usepackage{subcaption}

\usepackage{float}
\usepackage[linesnumbered,ruled,vlined]{algorithm2e}
\usepackage{placeins}
\usepackage{tikz}

\graphicspath{{./Figures/}}
\DeclareGraphicsExtensions{.pdf,.jpeg,.png,.jpg, .PDF, .JPEG, .PNG, .JPG}

\newcommand{\putsec}[2]{\section{#1}\label{sec:#2}}
\newcommand{\putsubsec}[2]{\subsection{#1}\label{sec:#2}}

\newcommand{\secref}[1]{Section~\ref{#1}}
\newcommand{\figref}[1]{Figure~\ref{#1}}
\newcommand{\tableref}[1]{Table~\ref{#1}}

\makeatletter
\def\blfootnote{\gdef\@thefnmark{}\@footnotetext}
\makeatother

\newcommand{\gnnaccel}{\texttt{\textsc{GNNerator}}}
\newcommand{\dense}{Dense Engine}
\newcommand{\graph}{Graph Engine}

\begin{document}

\title{GNNerator: A Hardware/Software Framework for Accelerating Graph Neural Networks
\thanks{This work was supported in part by C-BRIC, one of six centers
in JUMP, a Semiconductor Research Corporation (SRC) program
sponsored by DARPA}
}

\author{\IEEEauthorblockN{Jacob R. Stevens$^{1}$, Dipankar Das$^2$, Sasikanth Avancha$^2$, Bharat Kaul$^2$, Anand Raghunathan$^1$}
\IEEEauthorblockA{
\textit{Purdue University, West Lafayette$^1$}\\
\textit{Intel$^2$}\\
\{steven69, raghunathan\}@purdue.edu\\
\{dipankar.das, sasikanth.avancha, bharat.kaul\}@intel.com}
}

\maketitle

\begin{abstract}
Graph Neural Networks (GNNs) apply deep learning to inputs represented as graphs. They use fully-connected layers to extract features from the nodes/edges of a graph and aggregate these features using message passing between nodes, thereby combining two distinct computational patterns: dense, regular computations and sparse, irregular computations.
To address the computational challenges posed by GNNs, we propose \gnnaccel{}, an accelerator with heterogeneous compute engines optimized for these two patterns. 
Further, we propose feature-blocking, a novel GNN dataflow that beneficially trades off irregular memory accesses during aggregation for regular memory accesses during feature extraction. We show that \gnnaccel{} achieves speedups of \texttt{5.7-37x} over an NVIDIA RTX 2080-Ti, and \texttt{2.3x-3.8x} over HyGCN, a state-of-the-art GNN accelerator.
\end{abstract}


\putsec{Introduction}{intro}
Recently, graph neural networks (GNNs) have seen great success in achieving state-of-the-art results in a variety of applications, such as physics modeling \cite{interactions}, chemical synthesis \cite{deepgen}, and electronic design automation (EDA) \cite{circuit-gnn}. These GNNs consist of two main stages: the \emph{feature extraction} stage and the \emph{aggregation} stage. The feature extraction stage is typically a fully-connected network whose weights are shared across all nodes. In the aggregation stage, each node in the graph aggregates features from its neighbors into a new feature representation, resulting in sparse, random memory accesses.
Since each node is performing this aggregation independently, there is ample \emph{inter-node} parallelism. Further, in contrast to traditional graph-processing workloads such as PageRank where node features tend to be a few bytes, the features in GNNs can be thousands of bytes long. Due to this high dimensionality, there is also ample \emph{intra-node} parallelism, since the aggregation of each dimension can be done in parallel. Finally, the feature extraction and aggregation stages have a producer-consumer relationship, resulting in \emph{inter-stage} parallelism.

As a result of GNNs' unique computational characteristics,  traditional DNN accelerators are not well suited for GNNs. For example, DNN accelerators such as Eyeriss \cite{eyeriss} or Google's TPU \cite{tpu} are not optimized for the memory patterns present in the graph-based aggregation step, leading to poor utilization of on-chip resources. Similarly, graph analytics accelerators such as GraphP and Tesseract \cite{graphp, tesseract} are also ill-suited for GNNs, as these accelerators lack the compute resources for the dense, matrix-based operations found in the feature extraction stage. 

Due to the aforementioned challenges, there have been a few recent efforts to design GNN-specific accelerators. In particular, HyGCN \cite{hygcn} uses heterogeneous compute units for the feature extraction and aggregation stages, which are then tightly coupled together. This architecture allows HyGCN to address both major kernels described above. However, HyGCN has some drawbacks. Specifically, HyGCN fails to fully exploit inter-node parallelism, since it processes a single node at a time. Further, HyGCN only supports a dataflow in which the aggregation stage is the producer and the feature extractor is the consumer. This limits its applicability to workloads such as such as GraphsagePool, where this is not the case.

To overcome the aforementioned limitations, we propose \gnnaccel{} a  \underline{G}raph \underline{N}eural \underline{N}etwork accel\underline{erator} that exploits all three forms of parallelism found in GNNs: inter-node, intra-node, and inter-stage. \gnnaccel{} provides flexible, fine-grained control of how data flows between its dense engine and graph engine, supporting a wider variety of GNN topologies. We also propose a novel feature dimension-blocking dataflow. This  dataflow improves upon the dataflows used in previous GNN accelerators, reducing the overhead associated with processing the irregular, feature aggregation stage by exploiting the independent nature of the feature dimensions.
We evaluate \gnnaccel{} on a suite of nine GNN benchmarks and demonstrate that it outperforms a GPU baseline, as well as HyGCN.

In summary, we make the following contributions:

\begin{itemize}
\item We design \gnnaccel{}, a programmable graph neural network accelerator consisting of a Dense Engine and a Graph Engine, exploiting the inter-stage parallelism inherent in GNNs while allowing for flexibility in the producer-consumer relationship between them.
\item We provision the Dense and Graph Engines to exploit both the inter- and intra-node parallelism abundant in GNNs.
\item We propose a novel, feature-dimension blocking dataflow for GNNs and provide hardware support for this dataflow in  \gnnaccel{}.
\item We develop a simulation framework that implements the above proposals and use this framework to demonstrate the benefits of \gnnaccel{}. Our experiments indicate that \gnnaccel{} achieves an average speedup of \texttt{8x} over an NVIDIA 2080-Ti GPU and an average speedup of \texttt{3.15x} over HyGCN, a recent state-of-the-art GNN accelerator.

\end{itemize}
\putsec{Background}{background}

\putsubsec{Graph Neural Networks}{gnns}
The term Graph neural networks (GNNs) is used to collectively refer to a diverse family of networks. Many popular GNNs \cite{pinsage, gat} consist of two distinct stages: a \emph{feature extraction} stage, and an \emph{aggregation} stage. In the feature extraction stage, the node features are passed through one or more fully-connected linear layers. In the aggregation stage, a node aggregates feature vectors from its neighbors and then uses this aggregated feature vector to update its own feature.

These two stages are combined to make up a single GNN layer. Either stage may precede the other. A full GNN can then be constructed by stacking these layers. By stacking layers, a GNN incorporates information from nodes that are increasingly far away from the original node. For example, a single layer GNN only considers a node's neighbors, while a two layer GNN will consider nodes in the two-hop neighborhood. 


We provide a brief overview of two popular GNNs below in order to further illustrate this concept.

%

\noindent\textbf{Graphsage.} The Graphsage network \cite{graphsage} is an extremely popular example of GNNs. Graphsage applies the local weight sharing found in traditional convolutional neural networks to graphs such that every node (edge) feature shares the same weights. As shown in Equation~\eqref{eq:graphsage}, Graphsage first applies an aggregation stage, where the mean of the features of a node's neighbors is calculated. The aggregated node feature $\bar{z}$ is then passed through a linear layer to obtain the updated node feature, $h_u'$.

\begin{equation}
    \begin{gathered}
    \mathbf{\bar{z}} = \frac{1}{|N(u)|}\sum{\{\mathbf{h_v}|\forall v \in N(u) \cup u \}} \\
    \mathbf{h'} = \sigma(\mathbf{W} \cdot (\mathbf{\bar{z}} \cup \mathbf{h}))
    \end{gathered}
    \label{eq:graphsage}
\end{equation}

\noindent\textbf{GraphsagePool.} The GraphsagePool variant of Graphsage replaces the mean-based aggregator computation of $\bar{z'}$ with a symmetric, trainable aggregator, as described in Equation~\eqref{eq:graphsage-max}. Specifically, each node's feature is fed through a linear layer, defined by $W_{pool}$. The resulting features are then aggregated using an element-wise pooling operation (typically, max). Note that in this case, the feature extraction for $z$ is consumed by the aggregation for $\bar{z}$.

\begin{equation}
    \begin{gathered}
    \mathbf{z} = \sigma(\mathbf{W_{pool}} \cdot \mathbf{h}) \\
    \mathbf{\bar{z}} = max(\{\mathbf{z_v'} |\forall v \in N(u) \cup u \})
    \end{gathered}
    \label{eq:graphsage-max}
\end{equation}

\begin{figure}[h]
	\centering 
	\includegraphics[width=0.75\columnwidth]{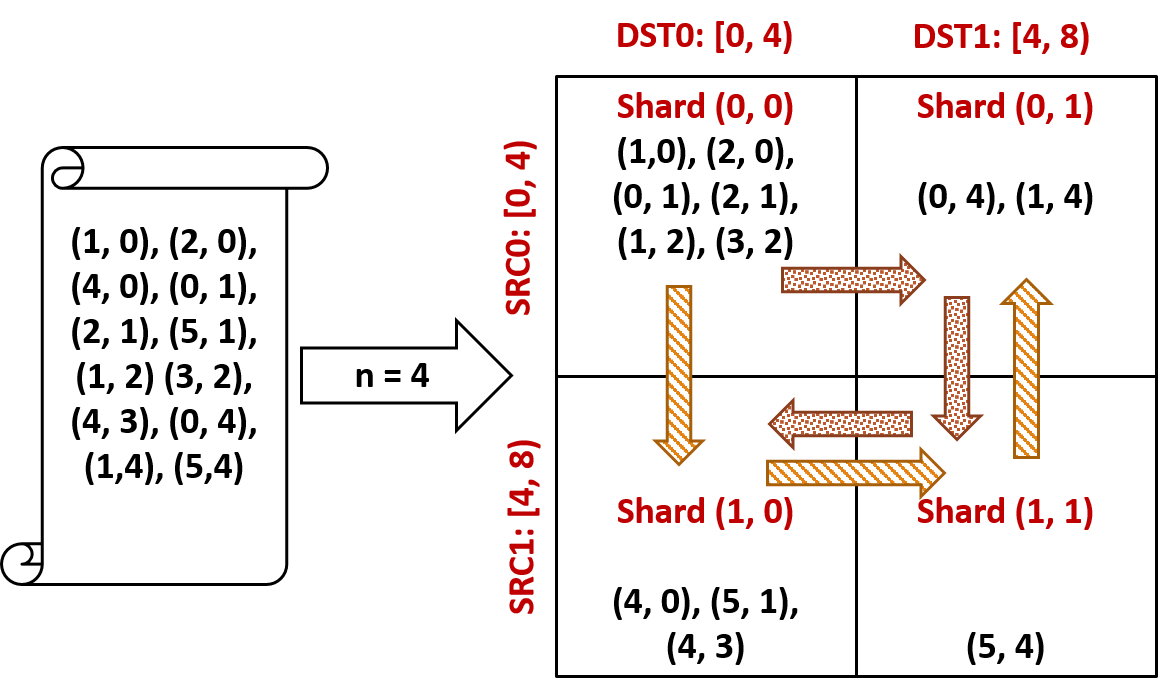}
	\caption[Visualization of a two-dimensional graph sharding algorithm]{A graph sharding algorithm divides an edge list into shards (subgraphs) which can then be processed in either a source-stationary (dotted arrow) or destination-stationary (slanted arrow) manner.}
	\label{fig:shard_visual}
\end{figure}

\vspace{-10pt}
\putsubsec{Graph Sharding}{sharding}
Many real-world graphs are too large to fit into a given level of the memory hierarchy, severely degrading the performance of graph processing algorithms, as it makes it extremely difficult to exploit any potential reuse. This is especially true when high-dimensional features are associated with the nodes/edges. To address this challenge, graphs are often broken into smaller pieces, such that each subgraph can fit in a given level of the memory hieraarchy. This process is referred to as \emph{graph sharding} \cite{gridgraph}. Similar to \cite{gridgraph}, we adopt a two-dimensional sharding paradigm, as depicted in \figref{fig:shard_visual}. In this paradigm, a graph's edge list is divided into shards such that each shard contains a maximum of $n^2$ edges, where $n$ is a tunable parameter that limits the number of source/destination nodes per shard.

\vspace*{-10pt}
\putsec{GNNerator Architecture}{gnn-arch}
\begin{figure}[b]
	\centering 
	\includegraphics[width=\columnwidth]{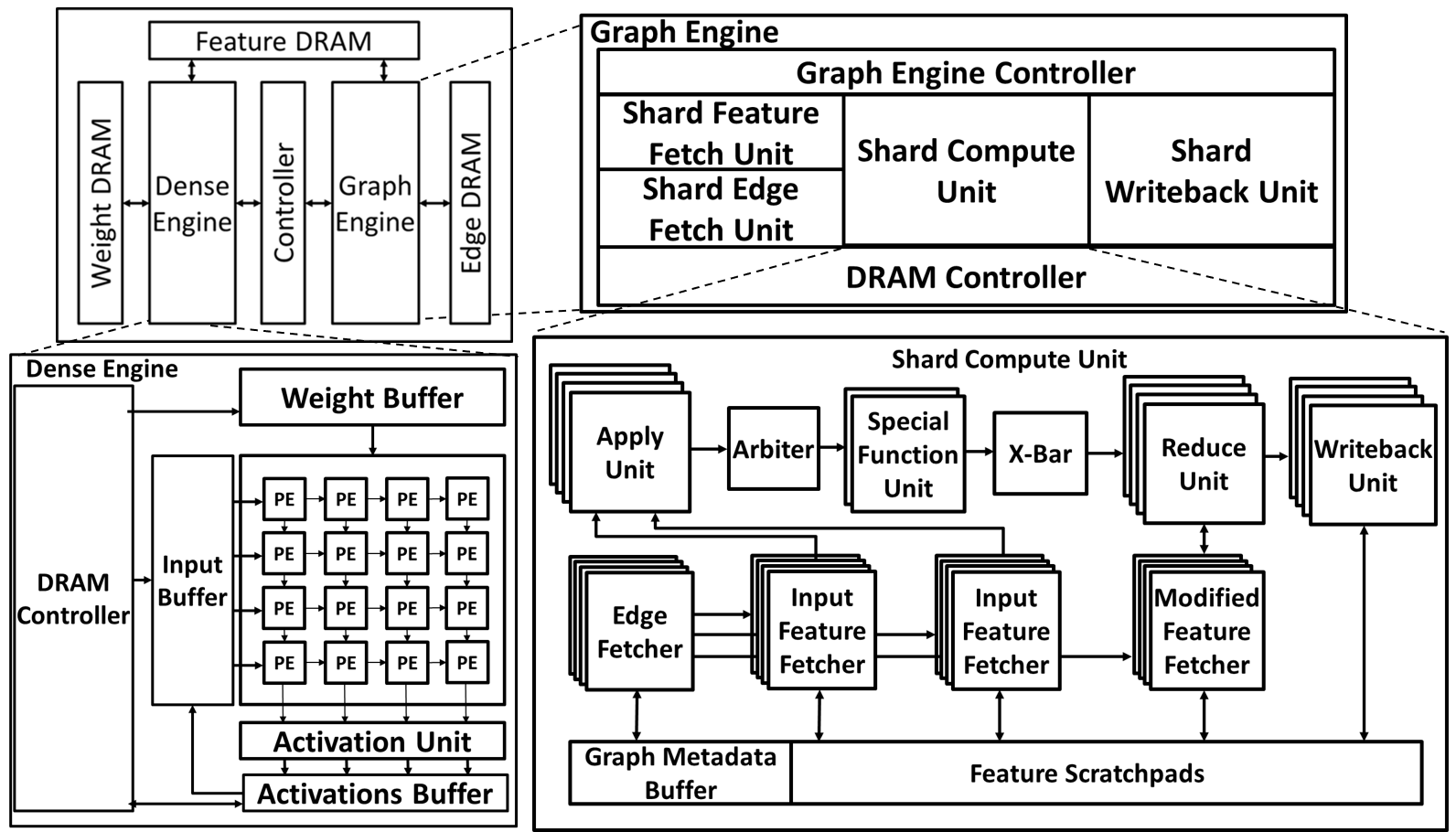}
	\caption[\gnnaccel{} architecture overview]{\gnnaccel{} (top left) consists of two heterogenous compute engines, a Dense Engine and a Graph Engine, that share feature storage.} 
	\label{fig:overview}
	\vspace{-0.1in}
\end{figure}

The \gnnaccel{} architecture, presented in Figure~\label{fig:overview}, consists of two different compute engines, the \dense{} and the \graph{}. The \dense{} is used for performing the dense, regular feature extraction steps of a GNN, while the \graph{} performs the graph-based aggregation steps. The interoperation of these engines is controlled by the \gnnaccel{} Controller.

\putsubsec{Dense Engine Overview}{dense-engine}
The \dense{} consists of a two-dimensional systolic array-based matrix multiplication unit, an activation unit, and input, weight, and activation on-chip double-buffered scratchpads, much like traditional DNN accelerators. The input and weight buffers feed the systolic array. The output of the two-dimensional systolic array is connected to a one-dimensional activation unit, which performs any required activation operations (\emph{e.g.}, ReLu). The results from the activation unit are stored in the output buffer, where they can either be transferred out to DRAM or to the input buffer to be reused as input to the systolic array. We note that, unlike HyGCN's combination engine, the Dense Engine has its own memory controller. This is necessary to support GNNs where the
Dense Engine needs to act as the producer. It also enables reloading of partial sums, which in turn facilitates the novel dataflow proposed in the following section.

In order to support fine-grain pipelining of the feature extraction and aggregation stages, the \dense{} also contains connections with the \graph{} that are used to communicate the current state of the respective computing engines.

\putsubsec{Graph Engine Overview}{graph-engine}
The \graph{} is tailored for the irregular nature of graph processing. It consists of four major units-- Shard Feature Fetch, Shard Edge Fetch,  Shard Compute, and Shard Writeback Units-- as well as a Graph Engine Controller that orchestrates the processing pipeline between the four units.

The Shard Edge Fetch and Feature Fetch Units work in parallel to load the edge data and feature data required to process a given graph shard from main memory to the on-chip scratchpads. After a shard is loaded, the Shard Compute Unit steps through its associated edges and performs the necessary computations. Finally, the Shard Writeback Unit stores output data from the on-chip scratchpads to main memory. As in the \dense{}, all of the on-chip buffers in the \graph{} are also double-buffered, enabling the pipelining of the above computations, such that the next shard is being prefetched while the current shard is being executed.


The Shard Compute Unit contains an Edge Fetcher, which steps through the edges associated with the current graph shard and distributes the required edge information (\emph{e.g}, source node ID) to the Feature Fetcher Units, as well as the Writeback Unit. These units use the edge information in order to generate read (write) accesses to the on-chip scratchpad. The fetch units feed the actual compute units: the Apply Unit, which performs binary operations, and the Reduce Unit, which performs an aggregation operation. Each of these compute units are vectorized in a single-instruction, multiple data (SIMD) manner  in order to exploit \emph{intra-node parellelism} across the different dimensions of a given node's feature, which are computed independently. 

In order to exploit \emph{inter-node parallelism}, the Shard Compute Unit contains multiple copies of the set of units described above, referred to collectively as a Graph Processing Element (GPE). 
Each GPE is assigned to a subset of the edges for a given graph shard, thereby processing multiple nodes at one time. 


\putsubsec{GNNerator Controller}{gnnerator-controller}
The GNNerator Controller coordinates the interaction between the \dense{} and the \graph{}. Crucially, our controller allows each engine to be either the producer or the consumer. This is important to efficiently support various GNN configurations, as in some networks the computation is a feature extraction followed by an aggregation and sometimes it is the other way around.

\noindent\textbf{Dense first.} If the feature extraction is first, then the \dense{} must run ahead of the \graph{}. Hence, the GNNerator Controller reads the state of the \dense{} and stalls the \graph{} until the source nodes for the current shard of the \graph{} have been processed by the \dense{}.

\noindent\textbf{Graph first.} If aggregation is first, then the \graph{} must run ahead of the \dense{}. Hence, the GNNerator Controller reads the state of the \graph{} and stalls the \dense{} until the \graph{} is done processing a set of destination nodes (\emph{i.e.}, the \graph{} has completed a full column of the shard grid).

\putsec{GNN Dataflows}{executionmodel}

In this section, we first describe the conventional dataflow used for executing graph neural networks, and then propose a novel feature dimension-blocking GNN dataflow, which we implement in \gnnaccel{}. The conventional GNN dataflow can be expressed using our proposed dataflow in Algorithm~\ref{alg:blockedalgo}, with $B$ set to the length of the node feature, $D$, thus eliminating the second loop (line 2).

\putsubsec{Conventional GNN Dataflow}{standard}
 In order to process a given sharded graph, \gnnaccel{} must step through the two-dimensional grid depicted in \figref{fig:shard_visual}. This can be done in a source-major or destination-major manner. This corresponds to the third and fourth loops in Algorithm~\ref{alg:blockedalgo} (lines 3-4); in this case, we are processing in a destination-major manner.

\SetKwInput{Input}{Input}
\SetKwInput{Output}{Output}
\SetKw{Let}{Let}
\SetKw{In}{in}
\SetKw{Range}{range}
\begin{algorithm}[tbhp]
\algsetup{linenosize=\scriptsize}
\small
\caption{Dimension-blocking Algorithm}
\label{alg:blockedalgo}
\Input{Sharded Graph $G$; Width/Height of Square Shard Grid $S$, Hidden Dimension Size $D$, Features $h$, Layers $L$}
\For{$l$ \In \Range($L$)}{
    \For{$blockD$ \In \Range($D/B$)}
        {
        \For{$dst$ \In \Range($S$)}{
            \For{$src$ \In \Range($S$)}{
                $Shard = G.Shards(src, dst)$ \\
                \For{$v$ \In \Range($Shard(src,dst).V$)}{
                    \For{$u$ \In \Range($v.U$)}
                    {
                        \For{$d$ \In \Range($B$)}
                        {
                            $dim = f(d, blockD)$ \\
                            $h_{agg}[v][dim]=\mathbf{Aggregate}(h_u[dim], h_v[dim])$
                        }
                    }
                }
            }
            $h'[dst][:] =$ \\
            $\mathbf{FeatureExtract}(h_{agg}[dst][blockD*B:(blockD+1)*B]$ $, h'[dst][:])$
        }
    }
    $h = h'$
}
\end{algorithm}

\begin{table}[htb]
\centering
\caption{Analytical description of shard dataflows}
\begin{tabular}{@{}lll@{}}
\toprule
        & Read Cost                & Write Cost    \\ \midrule
SRC Stationary &  $S*I+(S-1)*S-S+1$    & $S^2-S+1$ \\
DST Stationary  & $(S^2-S+1)*I$        & $S$         \\ \bottomrule
\end{tabular}
\label{table:rwcost}
\end{table}

When traversing in a source-major fashion (\emph{i.e.}, across a row of the shard grid), a set of \emph{source} vertices and their corresponding feature(s) are loaded on-chip and remain on-chip for the entire row. The destination vertices, however, must be written back and reloaded as we move from shard to shard. Conversely, when traversing in a destination-major fashion, a set of \emph{destination} vertices and their corresponding feature(s) are loaded on-chip and remain on-chip until they are done aggregating, while the source features must be reloaded as we move from shard to shard. The loading/storing of these features are performed by the Shard Feature Fetch Unit and the Shard Writeback Unit, respectively.

Assuming an S-pattern, we show the read and write costs associated with the two different orders in \tableref{table:rwcost}, where $S$ is the number of shards and $I$ is the maximum number of input features required to be on-chip at one time. With these costs, assuming equal costs for read and write transactions, we can analytically determine the best ordering.

Within each sub-graph shard, the edge list that defines the sub-graph must also be processed. The Edge Fetcher within the Shard Compute Unit steps through this edge list (corresponding to lines 6-7 in Algorithm~\ref{alg:blockedalgo}), and uses the list to program the Input Feature Fetcher Units, which perform the requisite loads. Finally, the nodes are aggregated (line 10) by the Apply and Reduce Units. Note that the Modified Feature Fetcher is used when reloading partial computed accumulations is required.
Finally, after aggregation, feature extraction (line 12) is performed by the Dense Engine. We note that in the traditional dataflow, wherein $B=D$, this is done in one step without the need to reload partial sums.

\putsubsec{Feature Dimension-blocking}{dimension-blocking}
We first note that the feature dimensions are treated \emph{independently} during  the  graph  processing  phase  of GNNs. With this insight in mind, we propose, and provide hardware support for, a novel feature-blocking dataflow (Algorithm~\ref{alg:blockedalgo}) wherein the graph processing computations are broken into two loops, such that only a block of dimensions is on-chip and processed at one time. A block is processed for every subshard before moving on to the next block of dimensions.

There are two main implications of the proposed dataflow. First, as shown in Table~\ref{table:rwcost}, the cost associated with read and writing the subshards is dependent on $S$, the number of shards. Thus, to minimize data transfers, we would like to maximize the number of nodes that can be held on-chip at one time -- that is, we would like the shards to be as large as possible. However, large shards are difficult to fit on-chip in the context of the traditational GNN dataflow, since each node or edge is associated with one or more features, each of which can be of a high dimension and must be held on-chip in the traditional dataflow. By contrast, in our proposed dataflow, only a subset of the dimensions are kept on chip at any time. Since each node requires less on-chip storage, more nodes can be stored on-chip, reducing $S$ in Table~\ref{table:rwcost}. Note, however, that the edge list is processed multiple times (\emph{i.e.}, lines 3-4), thus increasing the number of on-chip memory accesses. This is an overhead associated with feature-blocking, but it is favorably offset by the reduction in off-chip data transfers.

Second, since the traditional dataflow computes on entire features at once, the Dense Engine must load the weights for the entire feature, which are then shared across a relatively smaller number of nodes, for the reasons described above. In our dataflow, by contrast, the Dense Engine computes on more nodes, with fewer features per node at a time.  This increases reuse for the Dense Engine, improving efficiency and helping to mitigate the overheads of the new need to reload partial sums during feature extraction (line 12).

In effect, when comparing the conventional and proposed dimension-blocked dataflows, the irregular off-chip accesses of feature aggregation are reduced at the cost of additional regular accesses for feature extraction and additional on-chip accesses for processing the edges. As borne out by our results, this is a beneficial tradeoff. Moreover, the additional feature extraction accesses for partial sums are mitigated by the increased reuse enable by dimension-blocking. Finally, feature extraction is performed on only destination shards whereas aggregation depends on both source and destination shards, improving the benefits of dimension-blocking.

\putsec{Methodology}{methods}

\vspace{3pt}
\noindent\textbf{Simulation infrastructure.} We developed a cycle-level simulator as well as a prototype compiler and runtime for \gnnaccel{}. The cycle-level simulator was developed using the PyMTL3 framework \cite{mamba}. We implemented cycle-level models of all of the Graph Engine and GNNerator Controller components shown in \figref{fig:overview} and integrated the cycle-accurate SCALE-Sim simulator \cite{scalesim} for the Dense Engine.

\begin{table}[htb]
\caption{Summary of Graph Datasets}
\renewcommand{\arraystretch}{1.5}
\fontsize{9}{7.2}\selectfont
\begin{tabular}{lclcc}
\hline
\textbf{Dataset}  & \textbf{Vertices} & \textbf{Edges} & \textbf{Feature Dim.} & \textbf{Size}    \\ \hline
CORA     & 2708     & 10556 & 1433         & 15.6 MB \\
CITESEER & 3327     & 9104  & 3703         & 49 MB   \\
PUBMED   & 19717    & 88648 & 500          & 40.5 MB \\ \hline
\end{tabular}
\label{table:gnnerator-benchmarks}
\end{table}

\begin{table}[htb]
\caption{Summary of Graph Neural Networks}
\fontsize{9}{7.2}\selectfont
\renewcommand{\arraystretch}{1.5}
\begin{tabular}{ccc}
\hline
\textbf{Network}                & \textbf{Hidden Layers}    & \textbf{Hidden Dimension} \\ \hline
GCN \cite{gcn}                  & 1                         & 16 \\
Graphsage \cite{graphsage}      & 1                         & 16 \\
GraphsagePool \cite{graphsage}  & 1                         & 16 \\ \hline
\end{tabular}
\label{table:gnnerator-networks}
\end{table}

\vspace{3pt}
\noindent\textbf{Benchmarks.} In Table~\ref{table:gnnerator-benchmarks}, we summarize the input graph datasets used in our experiments. These datasets represent standard graph datasets used in GNNs. Note that most of the datasets cannot fit on-chip due to the large feature dimension sizes. We run each of these input graph datasets on the three graph neural network architectures outlined in Table~\ref{table:gnnerator-networks} using the Deep Graph Library (DGL) \cite{dgl} with the PyTorch backend.


\begin{table*}[tb]
\centering
\caption{Summary of Compute Platforms}
\begin{tabular}{cccc}
\hline
\multicolumn{1}{l}{}     & \textbf{RTX 2080 Ti}         & \textbf{GNNerator}                        & \textbf{HyGCN} \\ \hline
\textbf{Peak Compute}    & 13 TFLOPs            & 10 TFLOPs (2 for Graph, 8 for Dense)      & 9TFLOPs (1 for Graph, 8 for Dense)   \\
\textbf{On-chip Memory}  & 29.5 MiB             & 30 MiB (24 MiB Graph, 6 MiB Dense)        & 24 MiB \\
\textbf{Off-chip Memory} & 616 GB/s             & 256 GB/s                                  & 256 GB/s \\
\textbf{Area}           & 775 $mm^2$            & 14.5 $mm^2$                                  & 7.8 $mm^2$ \\ \hline
\end{tabular}
\label{table:gnnerator-platforms}
\end{table*}

\vspace{3pt}
\noindent\textbf{Platforms.} Table~\ref{table:gnnerator-platforms}
presents the configuration of \gnnaccel{} used in our evaluations, along with the relevant parameters for the baseline platforms: an NVIDIA RTX 2080 Ti GPU and HyGCN, a recently proposed state-of-the-art GNN accelerator. HyGCN was chosen as the baseline as it is the most similar design to \gnnaccel{} in terms of hardware resources (FLOPs and bandwidth).

\putsec{Evaluation}{gnn-evaluation}
\putsubsec{Performance}{pande}


\begin{figure}[b]
	\centering 
    \includegraphics[width=\columnwidth]{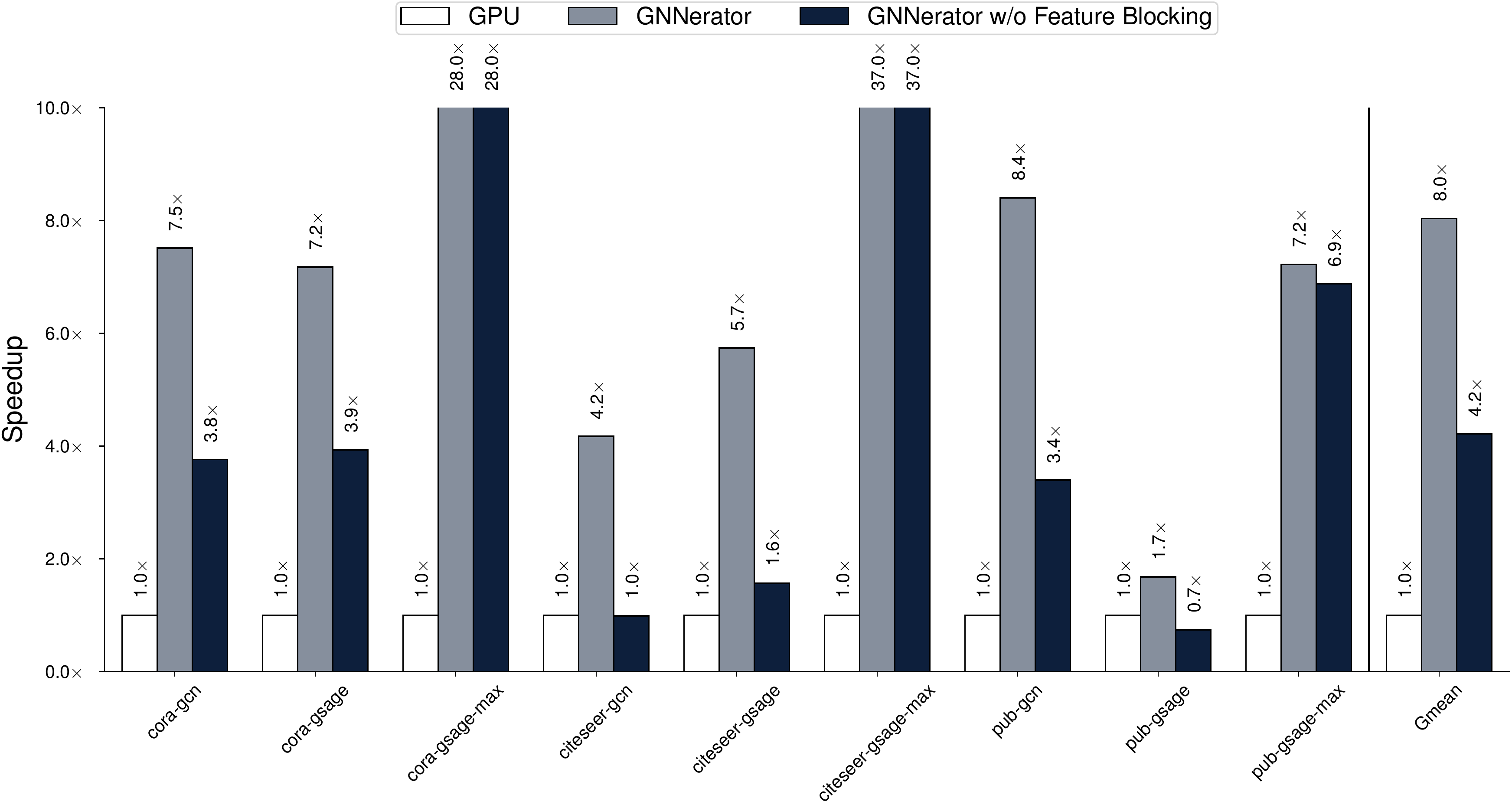}
	\vspace{-15pt}
	\caption[\gnnaccel{} speedup]{GNNerator achieves an \texttt{8x} speed up over the 2080-Ti baseline. Roughly half of this speedup results from the specialized architecture and the other half comes from the novel dimension blocking dataflow.}
	\label{fig:gnna_speedups}
	\vspace{-0.15in}
\end{figure}
\figref{fig:gnna_speedups} shows the normalized speedup with respect to the 2080-Ti GPU. We consider two variants of \gnnaccel{}: (i) the standard baseline \gnnaccel{} that uses the proposed novel dimension-blocking scheme described in \secref{sec:executionmodel} with $blockD$ set equal to the width of the Dense Engine (\emph{i.e.}, 64) and (ii) a variant that does not use dimension blocking. Without feature dimension-blocking, \gnnaccel{} demonstrates an average speed up of \texttt{4.2x} over the GPU baseline, while with feature dimension-blocking it has an average speedup of \texttt{8.0x}. Without feature blocking, \gnnaccel{}'s speedup comes from a variety of sources. First, as discussed in \secref{sec:gnn-arch}, \gnnaccel{} fully exploits all potential sources of parallelism in GNNs: inter-stage, inter-node, and intra-node parallelism. Further, unlike a GPU, the memory hierarchy of \gnnaccel{}'s Graph Engine is specialized for the unique needs of graph processing. For example, the memory width of the edge memories are sized such that no bandwidth is wasted due to mismatches between edge data size and memory width. \gnnaccel{}'s additional performance improvement through the use of dimension-blocking stems from two main sources. First, dimension-blocking allows for more nodes' features to be held on-chip, reducing the memory bottleneck of transferring these features on- and off-chip. Second, dimension-blocking reduces the amount of time the Dense Engine must wait for the Graph Engine to finish aggregating a node's neighborhood, as the Graph Engine only has to aggregate a small fraction of the dimensions before the Dense Engine can begin.



\begin{table}[]
\centering
\vspace{-0.2in}
\caption{Speedups of \gnnaccel{} over HyGCN for GCN}
\begin{tabular}{cccc}
\hline
\textbf{}               & \textbf{Cora} & \textbf{Citeseer} & \textbf{Pubmed} \\ \hline
\textbf{GNNerator w/o blocking} & 1.8x            & 0.8x              & 1.0x            \\
\textbf{GNNerator}      & 3.8x          & 3.2x              & 2.3x            \\ \hline
\end{tabular}
\label{table:hygcn}
\end{table}

\noindent\textbf{HyGCN Comparison.} Table~\ref{table:hygcn} compares \gnnaccel{} to the recently proposed state-of-the-art GNN accelerator HyGCN using their results described in \cite{hygcn}. We chose HyGCN as the comparison platform, as it is  closest to \gnnaccel{} in terms of hardware resources. As demonstrated in the table, without feature dimension-blocking, \gnnaccel{} and HyGCN are quite similar in performance, with comparable performance improvements over the GPU baseline. We note that HyGCN uses a sparsity elimination optimization which is particularly effective for the Citeseer dataset (roughly, \texttt{1.1x} for Cora/Pubmed, \texttt{3x} for Citeseer). This optimization is orthogonal to our work and can be added to \gnnaccel{}. When utilizing dimension-blocking, however, \gnnaccel{} consistently and considerably outperforms HyGCN, with an average speedup of \texttt{3.15x}.

\begin{figure}[tb]
	\centering 
    \includegraphics[width=0.8\columnwidth]{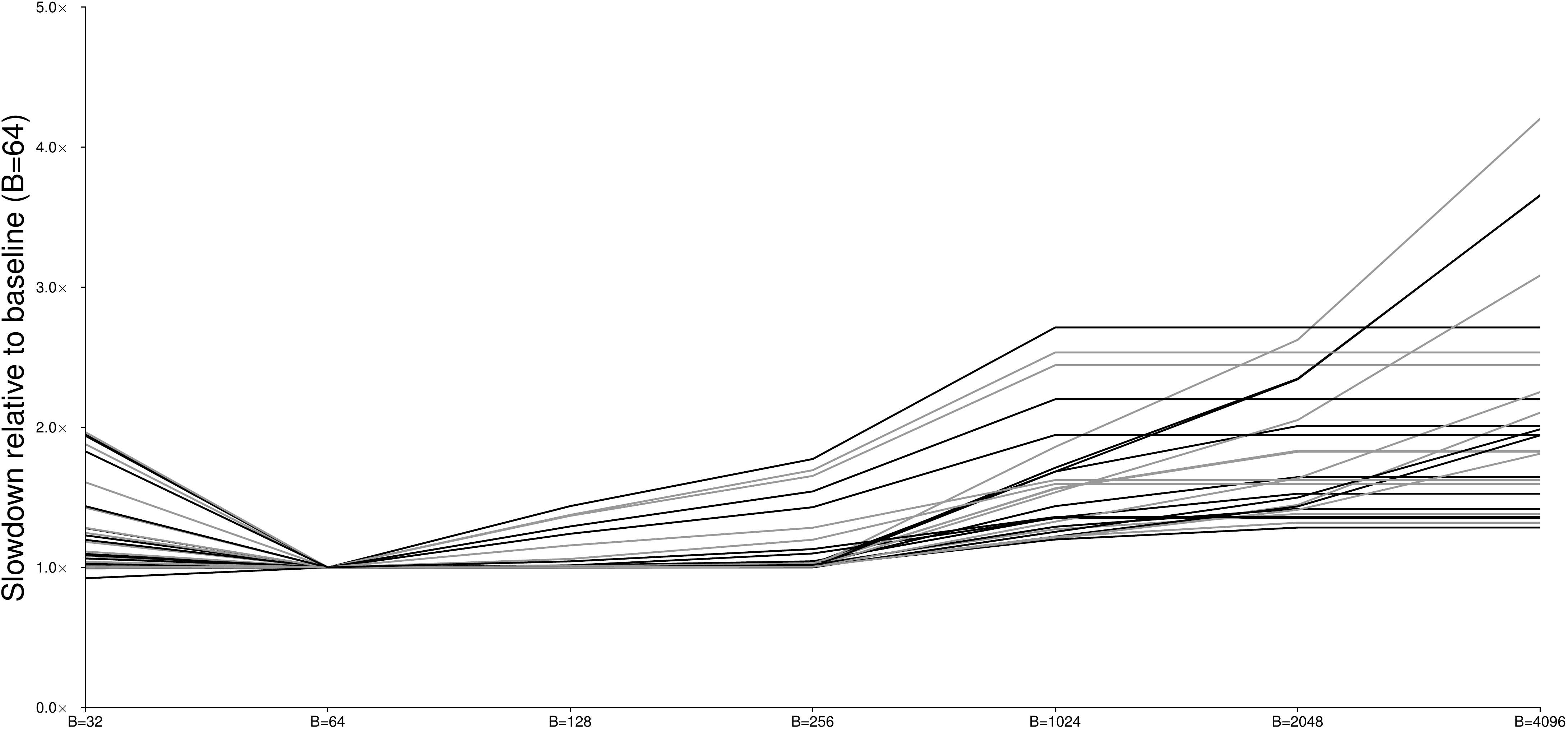}
	\vspace{-3pt}
	\caption[Feature-blocking ablation]{Generally, a smaller feature block size $B$ is desirable. However, performance suffers if $B$ is less than the width of the systolic array of the Dense Engine (\emph{i.e}, $B=32$)}
	\label{fig:blocking_ablation}
	\vspace{-0.15in} 
\end{figure}

\noindent\textbf{Feature-blocking Ablation.} A key question regarding our novel feature-blocking is the optimal size of the feature block, $B$-- that is, how many dimensions of the feature will be kept on-chip at once. To explore this, we performed a sweep of this parameter, while running a large number of various networks and datasets. As seen in \figref{fig:blocking_ablation}, a smaller $B$ is desirable. However, there is a lower limit to this. If the block size is set to the smallest possible value (\emph{i.e.}, the width of the Graph Engine lanes), the performance suffers. This is because the block size is then less than the width of the Dense Engine systolic array of sixty-four, leading to under-utilization of the engine's resources. 

\putsubsec{Scaling GNNerator}{scaling}

Finally, we consider the scenario of scaling \gnnaccel{} to larger chip sizes and examine the question of where to invest the additional hardware resources in order to maximize the performance return on that investment. We present three possible versions of a ``next-generation" \gnnaccel{}. One version doubles the amount of on-chip memory in the Graph Engine, allowing for more larger shards to be held on-chip. Another version doubles both the height and width of the Dense Engine, increasing the compute available for the linear layers. The final version doubles the bandwidth available for the shared feature memory DRAM.

\begin{figure}[tb]
	\centering 
	\includegraphics[width=\columnwidth]{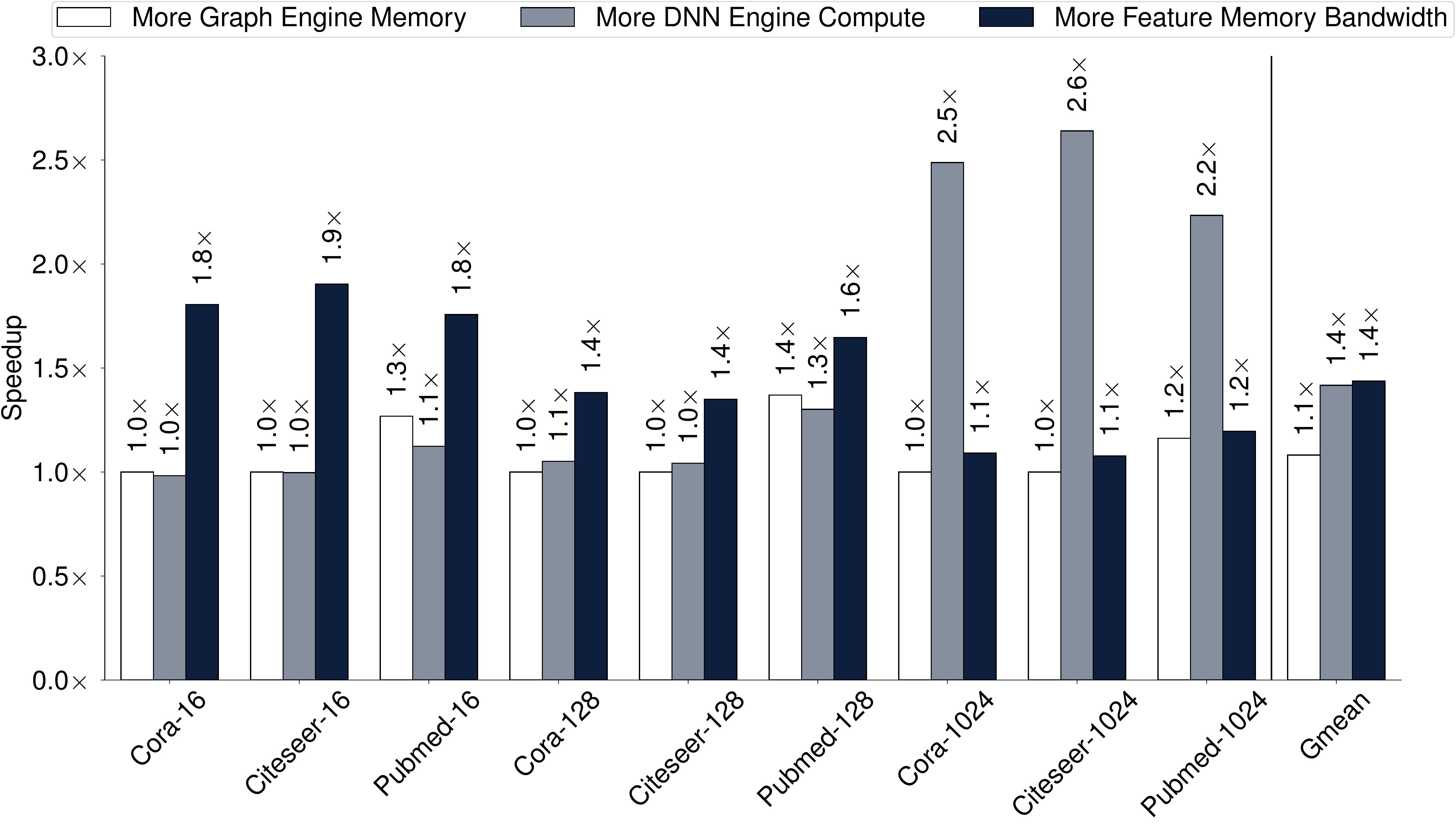}
	\vspace{-15pt}
	\caption[Scaling up \gnnaccel{}]{Adding feature memory bandwidth tend to improve performance for networks with smaller hidden dimension size while more DNN Engine compute results in the largest speedups for networks with large hidden size.}
	\label{fig:bigger_gnnas}
	\vspace{-0.2in} 
\end{figure}

We find that the best investment depends on the target networks. For networks with smaller hidden dimension sizes, increasing the feature memory bandwidth provides the highest return. At larger hidden sizes, on the other hand, increasing the size of the Dense Engine results in the largest speedups. This is driven primarily by large speedups for the larger hidden dimension sizes, as seen in \figref{fig:bigger_gnnas}.
\putsec{Discussion and Related Work}{related}




Hardware acceleration of neural networks has been a very active topic in the past decade. However, very few efforts directly target graph neural networks. The closest effort to \gnnaccel{} is HyGCN \cite{hygcn}, which proposes two heterogeneous compute engines, one optimized for the feature extraction stage and one for the aggregation stage. 
Our work differs from HyGCN in a few major ways. First, unlike in our work, HyGCN only exploits intra-node parallelism, processing a single node's entire feature across all cores before moving on to the next node. In contrast, \gnnaccel{} exploits both intra-node and inter-node parallelism. Second, \gnnaccel{}'s dual engine architecture is more flexible. For example, in \gnnaccel{}, the Graph Engine can be either the producer or the consumer-- in HyGCN, the Aggregation Engine is the only producer. This directly limits the applicability of HyGCN in workloads where the Dense Engine is the producer, such as GraphsagePool. Finally, HyGCN does not contain the architectural support necessary for  feature blocking.

Due to the wide variety in hardware resources used and benchmarks evaluated, it is difficult to compare \gnnaccel{} to other recent GNN proposals such as GNNA \cite{dacgnna} and EnGN \cite{engn}. However, these architectures rely on the traditional GNN dataflow and can therefore benefit from our proposed feature blocking-based dataflow with suitable enhancements.



\putsec{Conclusion}{gnn-conclusion}
GNNs are a promising new area of machine learning that aim to bring the success of deep learning from Euclidean domains to graph-based inputs. In this work, we detail the limitations of current hardware in efficiently realizing GNNs and propose \gnnaccel{}, a specialized hardware accelerator for GNNs that is able to exploit the abundant intra- and inter-node parallelism inherent in GNNs. \gnnaccel{} utilizes feature dimension-blocking, a novel GNN dataflow that allows for processing more nodes in a graph on-chip at one time. We evaluate \gnnaccel{} on a suite of benchmarks and demonstrate significant performance benefits over GPUs, as well as a recently proposed GNN accelerator. 




\bibliographystyle{IEEEtran}
\bibliography{refs}

\end{document}